# PROPERTIES OF CARRY VALUE TRANSFORMATION

**[1]Suryakanta Pal, [2]Sudhakar Sahoo and [3]Birendra Kumar Nayak**

[1]Silicon Institute of Technology, Silicon Hills, Patia, Bhubaneswar-751024

Email:-surya_tuna@yahoo.co.in

[2]Institute of Mathematics and Applications, Andharua, Bhubaneswar -751003

Email:-sudhakar.sahoo@gmail.com

[3]P.G. Department of Mathematics, Utkal University, Bhubaneswar-751004

Email:-bknatuu@yahoo.co.uk

**Abstract:-**The notion of Carry Value Transformation (CVT) is a model of Discrete Deterministic Dynamical System. In this paper, we have studied some interesting properties of CVT and proved that (1) the addition of any two non-negative integers is same as the sum of their CVT and XOR values. (2) While performing the repeated addition of CVT and XOR of two non-negative integers "a" and "b" (where a≥b), the number of iterations required to get either CVT=0 or XOR=0 is at most the length of "a" when both are expressed as binary strings. A similar process of addition of Modified Carry Value Transformation (MCVT) and XOR requires a maximum of two iterations for MCVT to be zero. (3) An equivalence relation is defined in the set $Z \times Z$ which divides the CV table into disjoint equivalence classes.

**Keywords:-Carry Value Transformation (CVT) and Modified Carry Value Transformation (MCVT).**

## 1. Introduction

The notion of transformation is very important in Mathematics. Accordingly, in literature, one finds many kinds of transformations with interesting properties. Carry Value Transformations (CVTs) and Modified Carry Value Transformations (MCVTs) are two challenging transformations which currently have assumed much significance because of its applications in fractal formation [1], designing new hardware circuits for arithmetic operations [2] etc. Similar kind of transformations such as Extreme Value Transformation (EVT) [3], 2-Variable Boolean Operation (2-VBO) [6], Integral Value Transformation (IVT) [7] are also applicable in pattern formations [3, 6], solving Round Rabin Tournaments problem [5], Collatz like functions [7] etc. All these applications in diversified domain motivated us to study the mathematical properties of these kinds of transformations.

The hardware circuit for arithmetic operations as designed in [2] is based on a result that after finite number of iterations, either CVT of the two non-negative integers is equal to 0 or their XOR value is equal to 0. But no mathematical proof regarding this result was discussed in [2]. This important result has been proved in this paper. We have also found other interesting results of CVT and MCVT. Section 2 provides the basic concepts of CVT, MCVT and XOR earlier defined in [1, 2]. In section 3, we have proved that addition of any two non-negative integers is same as addition of their CVT and their XOR values. This result is also shown to be true for any base of the number system. In section 4, we have proved that in a successive addition of CVT and XOR of any two non-negative integers, the maximum number of iterations required to get either CVT=0 or XOR=0 is equal to the length of the bigger integer. Also in the same section we have proved that MCVT of any two non-negative integers = 0 requires a maximum of two iterations. In section 5, we have formed an equivalence relation using the concept of CVT and the equivalence classes obtained due to it.

## 2. Definitions of CVT and MCVT in binary number system

Let "a" and "b" be decimal representations of the binary strings $(a_n, a_{n-1},....,a_1)_2$ and $(b_n, b_{n-1},....,b_1)_2$ respectively and $B_2^n$ be the set of strings of length $n$ on the set $B_2 = \{0,1\}$. In binary number system, CVT as defined in [1] is a mapping $CVT: B_2^n \times B_2^n \rightarrow B_2^n \times \{0\}$ defined by $CVT(a,b) = (a_n \wedge b_n, a_{n-1} \wedge b_{n-1},.........,a_1 \wedge b_1, 0)_2$ where as MCVT in [1] is a mapping $MCVT: B_2^n \times B_2^n \rightarrow B_2^n$ defined by $MCVT(a,b) = (a_n \wedge b_n, a_{n-1} \wedge b_{n-1},.........,a_1 \wedge b_1)_2$ That is to find out CVT, we perform the bit wise XOR operation of the operands to get a string of sum-bits (**ignoring the carry-in**) and simultaneously the bit wise ANDing of the operands to get a string of carry-bits, the latter string is padded with a '0' on the right is called the CVT of these operands as shown in fig 1.

$$
\begin{array}{llll}
carry\ value = a_n \wedge b_n & a_{n-1} \wedge b_{n-1}............a_1 \wedge b_1 & 0 \\
\hline
a = & a_n & a_{n-1}.............a_1 \\
b = & b_n & b_{n-1}.............b_1 \\
\hline
a \oplus b = & a_n \oplus b_n & a_{n-1} \oplus b_{n-1}.........a_1 \oplus b_1
\end{array}
$$

Fig 1: Carry generated in $i^{th}$ column counted from LSB is saved in $(i+1)^{th}$ column

For example, suppose we want the CVT of the numbers $(23)_{10} \equiv (10111)_2$ and $(27)_{10} \equiv (11011)_2$. The carry value is computed as follows.

$$
\begin{array}{l}
\underline{\text{Carry: } 1\ 0\ 0\ 1\ 1\ 0} \\
\underline{\text{Augend: } 1\ 0\ 1\ 1\ 1} \\
\underline{\text{Addend: } 1\ 1\ 0\ 1\ 1} \\
\text{XOR: } \quad 0\ 1\ 1\ 0\ 0
\end{array}
$$

Fig 2: Carry generated in $i^{th}$ column is saved in $(i+1)^{th}$ column for i=0, 1, 2,...,n.

Thus, CVT (23, 27) = CVT (10111, 11011) = $(100110)_2 = (38)_{10}$. It may be noted that in any number system, CVT is a mapping from $Z \times Z$ to $Z$ where $Z$ is set of non-negative integers.

### 2.1 Extensions of CVT, MCVT and XOR operations for arbitrary number system

For any number system in base $\beta$, CVT of any two non-negative integers $a = (a_n, a_{n-1},......,a_1)_\beta$ and $b = (b_n, b_{n-1},...,b_1)_\beta$ is defined by an integer $c = (c_1 c_2 .............c_n 0)_\beta$ where

$$
c_i = \begin{cases} 1, \text{ if } a_i + b_i \geq \beta \\ 0, \text{ if } a_i + b_i < \beta \end{cases}
$$

for i =1, 2, 3..., n. Similarly, MCVT of a and b in base $\beta$ is the CVT value $c = (c_1 c_2 .............c_n)_\beta$ except the padding bit 0 in the least significant bit position. That is $CVT(a,b) = \beta \times MCVT(a,b)$ and the definition of XOR operation in binary number system can be extended for any number system in base $\beta$ as $a \oplus b = ((a_n + b_n) \mod \beta, (a_{n-1} + b_{n-1}) \mod \beta,.........,(a_1 + b_1) \mod \beta)$ where + is the usual addition in decimal number system.

For example in ternary number system, CVT(466, 458) = CVT(122021, 121222) = $(110110)_3 = 336$, MCVT(466 , 458) = MCVT(122021, 121222) = $(11011)_3 = 168$, XOR(466, 458)= XOR(122021, 121222) = $(210210)_3 = 588$

## 3. Properties of CVT and XOR

We have observed in last example that CVT (23, 27)=38 and XOR(23,27)=12. Now 23+27 =38+12 .i.e. 23+27=CVT (23, 27) + $(23 \oplus 27)$ In general, we prove the following.

**Theorem 1:** a +b=CVT (a, b) + $(a \oplus b)$ where a and b are any two non-negative integers.

**Proof:** Suppose a=$a_n a_{n-1}........a_{k-1} a_k a_{k+1}........a_2 a_1$ and b=$b_n b_{n-1}.........b_{k-1} b_k b_{k+1}........b_2 b_1$ are the binary representations of a and b both expressed using n bits. Then CVT (a, b)= $c_n c_{n-1} c_{n-2}.............c_1 0$ for i= 1, 2,.........,n.We will prove that sum of the contribution of $a_k$ and $b_k$ in a+b is same as the sum of the contribution of $c_k$ and $a_k \oplus b_k$ in CVT(a, b) +$(a \oplus b)$ where k =1,2 ,3..........,n . The place values of $a_k$ and $b_k$ in a and b are $a_k \times 2^{k-1}$ and $b_k \times 2^{k-1}$ respectively. So the total contributions of both $a_k$ and $b_k$ in a +b is $(a_k + b_k) 2^{k-1}$. The binary variable $a_k$ and $b_k$ can have four choices and their place values are shown in table 1.

Table 1: Shows the contributions in calculating the sum in different cases

| $a_k$ | $b_k$ | Sum of contributions of $a_k$ and $b_k$ in a +b | $c_k = a_k \wedge b_k$ | Contribution of $c_k$ in CVT(a, b) | Contribution of $a_k \oplus b_k$ in $(a \oplus b)$ | Sum of contributions of $c_k$ and $a_k \oplus b_k$ in CVT(a, b) +$(a \oplus b)$ |
|---|---|---|---|---|---|---|
| 0 | 1 | $2^{k-1}$ | 0 | 0 | $2^{k-1}$ | $2^{k-1}$ |
| 1 | 0 | $2^{k-1}$ | 0 | 0 | $2^{k-1}$ | $2^{k-1}$ |
| 0 | 0 | 0 | 0 | 0 | 0 | 0 |
| 1 | 1 | $2^k$ | 1 | $2^k$ | 0 | $2^k$ |

From third column and seventh column, it can be verified that the total contribution of $a_k$ and $b_k$ in a+b is same as the sum of the contribution of $c_k$ and $a_k \oplus b_k$ in CVT(a, b) +$(a \oplus b)$ for k=1,2 ,...,n. Therefore a +b=CVT (a, b) +$(a \oplus b)$. Hence proved.

## A) General Proof for an arbitrary number system

Let a $= \sum_{k=1}^{n} a_k \times \beta^{k-1}$ and b$= \sum_{k=1}^{n} b_k \times \beta^{k-1}$ be two numbers from a number system with base $\beta$.

Let CVT (a, b)= $c_n c_{n-1}................c_1 0$ . We will prove that sum of the contribution of $a_k$ and $b_k$ in a+b is same as the sum of the contribution of $c_k$ and $a_k \oplus b_k$ in CVT (a, b) +$(a \oplus b)$ where k= 1, 2, 3,......., n.

Note that the individual place values of $a_k$ and $b_k$ in a and b are $a_k \times \beta^{k-1}$ and $b_k \times \beta^{k-1}$ respectively. So the total contributions for both $a_k$ and $b_k$ in a +b is $(a_k + b_k) \beta^{k-1}$. Two cases arises. Case:-1 $a_k + b_k < \beta$ and Case:-2 $a_k + b_k \geq \beta$

Table 2: Shows the contributions in calculating the sum for two possible cases

| Cases | Conditions | Sum of contributions of $a_k$ and $b_k$ in a +b | $c_k$ | Contribution of $c_k$ in CVT(a, b) | Contribution of $a_k \oplus b_k$ in $(a \oplus b)$ | Sum of contributions of $c_k$ and $a_k \oplus b_k$ in CVT(a, b) +$(a \oplus b)$ |
|---|---|---|---|---|---|---|
| Case:-1 | $a_k + b_k < \beta$ | $(a_k + b_k) \beta^{k-1}$ | 0 | 0 | $(a_k + b_k) \beta^{k-1}$ | $(a_k + b_k) \beta^{k-1}$ |
| Case:-2 | $a_k + b_k \geq \beta$ | $(a_k + b_k) \beta^{k-1}$ | 1 | $\beta^k$ | $(a_k + b_k) \beta^{k-1} - \beta^k$ | $(a_k + b_k) \beta^{k-1}$ |

From third column and last column , we observe that sum of the contribution of $a_k$ and $b_k$ in a+b is same as the sum of the contribution of $c_k$ and $a_k \oplus b_k$ in CVT(a, b) $+ (a \oplus b)$ where k=1,2, 3,4 ,…,n. Therefore, a +b=CVT (a, b) $+ (a \oplus b)$.

# 4. Convergence behavior of CVT and MCVT

## 4.1 Convergence of CVT

Let $f : Z \times Z \rightarrow Z \times Z$ be defined as $f(a,b) = $ (CVT(a ,b), (a $\oplus$ b)) for all (a, b) $\in$ $Z \times Z$. Consider the iterative scheme $(x_{n+1}, y_{n+1}) = f(x_n, y_n)$, n=0, 1, 2, 3,… In this section, we will prove an important theorem which states that the sequence generated by the iterative scheme $(x_{n+1}, y_{n+1}) = f(x_n, y_n)$, n=0,1,2,3,…..converges to (0 , $x_0 + y_0$). The convergence behavior of CVT and XOR values of different order pairs are shown in table 3.

Table 3: Generated sequences of CVT and XOR values

| Initial guess $(x_0 , y_0)$ | Generated sequences $(x_{n+1}, y_{n+1})$ |
|---|---|
| (1, 8) | (0, 9) |
| (12, 10) | (16, 6), (0, 22) |
| (12, 25) | (16 ,21), (32, 5), (0, 37) |
| (14, 22) | (12, 24), (16 , 20), (32, 4), (0, 36) |
| (21, 27) | (34 , 14), (4, 44), (8, 40), (16, 32), (0, 48) |
| (27, 5) | (2, 30), (4, 28), (8, 24), (16, 16), (32, 0), (0, 32) |
| (127, 65) | (130, 62), (4, 188), (8, 184), (16, 176), (32, 160), (64, 128), (0,192) |

The sequences generated from the ordered pair (127, 65) in table 3 may be interpreted as 127 + 65 = 130+62=4+188= 8+184=16+176=32+160=64+128=0+192.These generated sequences are named as the orbit of the order pair (127, 65).

**Observations:**
**(1)** CVT = 0 in i-th iteration $\Leftrightarrow$ CVT and XOR obtained in (i-1)[th] iteration have no "1" in the same position.
**(2)** If two numbers c and d are different in all positions in their binary representations in any iteration , then CVT( c ,d)=0 in that iteration .But converse is not true .
**(3)** a=b $\Leftrightarrow$ $a \oplus$ b=0
**(4)** If XOR value =0 in any iteration, then new CVT =0 in next iteration.
**(5)** According to definition of CVT, $CVT : B_2^n \times B_2^n \rightarrow B_2^n \times \{0\}$ where B₂= {0, 1}i.e. for any two bit n-bit numbers , CVT will be of at most (n+1) bits. In 2[nd] iteration, we add this CVT with XOR obtained in 1[st] iteration. Since CVT is of (n+1) bits so in 2[nd] iteration, new CVT will be of at most (n+2) bits as per definition. In 3[rd] iteration, CVT will be of at most (n+3) bits as per definition and so on. But it is not true which is clear from the next proof.

**Lemma 1:** If a and b are of maximum n binary bits, all the CVTs and XORs will be of at most (n+1) bits in all iterations while performing the repeated sum of CVT and XOR.

**Proof:** Let c and d be two numbers to be added in an iteration while performing the repeated sum of CVT and XOR . Suppose the number of (valid) bits in CVT(c, d) $\geq$ n+2 . Zero the number of (valid) bits in CVT(c, d) $\geq$ n+2 (rejecting the zeros in the left of the first nonzero bit) in an iteration. The smallest number using (n+2) bits or more is 100………0= $1 \times 2^{n+1} = 2^{n+1}$.
So CVT (c,d) $\geq$ $2^{n+1}$ $\Rightarrow$ CVT ( c ,d) $+ (c \oplus d)$ $\geq$ $2^{n+1}$ $\Rightarrow$ c +d $\geq$ $2^{n+1}$ [Using Theorem-1]

$\Rightarrow a+b \geq 2^{n+1}$      $[\because c+d=a+b]$.............................(1)

If a and b are maximum of n bits numbers , then maximum value of a and b is 111..........11

$= 1 \times 2^{n-1} + 1 \times 2^{n-2} + .....................+ 1 \times 2^1 + 1 \times 2^0 = 1+2+4+8+............. +2^{n-2} + 2^{n-1}$

Maximum value of a+b is $2(1+2+4+................ +2^{n-2} + 2^{n-1}) = \dfrac{2\left(2^n - 1\right)}{2-1} = 2^{n+1} - 2$

$\Rightarrow a + b \leq 2^{n+1} - 2$ ..............................(2)

From equation (1) and (2), we get $2^{n+1} \leq a+b \leq 2^{n+1} - 2$ which is absurd. Thus our assumption was wrong and hence all CVTs will be of at most (n+1) bits in every iteration.

Same logic can be applied to XOR operation also, i.e. if we write CVT in place of XOR in above proof, we also get an absurd result for XOR. Therefore all XOR operations are of at most (n+1)bits in every iteration.

**Lemma 2:** If there is a "0" in CVT at k$^{th}$ position(counted from left) in any iteration, then there must be one "0" in (k+1)$^{th}$ position in next iteration while forming the subsequent CVTs, but number of zeros in a CVT increases by at least one in each iteration.

**Proof:** Suppose a CVT contains 0 at k-th position in any iteration. In next iteration, this 0 will be added to either 0 or 1 of XOR obtained in the previous iteration. When we form CVT, (k+1)$^{th}$ position of CVT will be either $0 \wedge 1 = 0$ or $0 \wedge 0 = 0$. Thus we get a 0 in (k+1)$^{th}$ position of the newly formed CVT. Thus once a "0" appears in a CVT in any iteration, then a "0" appears in all subsequent CVT's in all subsequent iterations, but the position will be shifted by one in each iteration. By definition of CVT, one artificial zero is added to the rightmost position in each iteration. So number of zero increases by at least one in a CVT in each iteration.

**Lemma 3:** If a and b are of maximum n binary bits, then the number of iterations required to get CVT=0 is at most (n+1).

**Proof:** By Lemma:-1, all CVTs will be of at most (n+1) bits in all iterations.
By Lemma:- 2, once a "0" appears in a CVT in any iteration , then this zero will appear in all the subsequent CVT's in all subsequent iterations, but the position will be shifted by one in each iteration.
Also the number of zero in CVT increases by at least one in each iteration,
The (n+1) bits in CVT will be converted to (n+1) zeros in at most (n+1)-iterations.

**Note:** If a and b are of maximum n binary bits and Hamming distance between a and b is n , then CVT=0 in one iteration. Otherwise, if Hamming distance between two selected numbers is k for k < n, then number of iterations required to get CVT=0 is at most (k+2).

**Lemma 4:** If a and b are of maximum n binary bits and CVT=0 in (n+1)$^{th}$ iteration, then XOR=0 in n$^{th}$ iteration.

**Proof:** Let us assume that CVT=0 in the (n+1)$^{th}$ iteration and suppose XOR $\neq 0$ in the n$^{th}$ iteration. Then at least one bit of the XOR in n$^{th}$ iteration must be "1". In the k$^{th}$ iteration (where k=1 ,2 ,3,.........or (n-1)), XOR bit must be 1 and the corresponding carry bit must be 1 which is impossible. So our assumption was wrong. Thus XOR=0 in the n$^{th}$ iteration. Hence proved.

Combining Lemma:-3 and Lemma:-4, we have proved the following theorem.

**Theorem 2:** Let $f : Z \times Z \rightarrow Z \times Z$ be defined as $f(a,b) = $ (CVT(a, b), (a $\oplus$ b)). Then the iterative scheme $(x_{n+1}, y_{n+1}) = f(x_n, y_n)$, n=0, 1, 2, 3,... converges to $(0, x_0 + y_0)$ for any initial choice $(x_0, y_0) \in Z \times Z$. Further, for any non-negative integers " $x_0$ " and " $y_0$ " (where $x_0 \geq y_0$), the number of iterations required to get either CVT=0 or XOR=0 is at most the length of " $x_0$ " when expressed as a binary string.

## 4.2 Convergence of MCVT

The following lemma gives the number of iterations required for MCVT=0.

**Theorem 3:** The procedure of calculating the MCVT and XOR values of any two non-negative integers requires a maximum of two iterations to get their MCVT=0.

**Proof:** Let a= $a_1 a_2 ............ a_n$ and b= $b_1 b_2 ............. b_n$ be two n-bits number. In the first iteration, we get MCVT (a, b) and a $\oplus$ b.

Let x = MCVT (a, b) = $\left(a_n \wedge b_n, a_{n-1} \wedge b_{n-1}, ...................................................., a_1 \wedge b_1\right)$ and y = a $\oplus$ b= $\left(a_n \oplus b_n, a_{n-1} \oplus b_{n-1}, ...................................................., a_1 \oplus b_1\right)$. Then in the second iteration, we get MCVT (x, y) and (x $\oplus$ y) . We will show that MCVT (x, y)=0.

Table 4: Showing calculation of MCVT

| $a_i$ | $b_i$ | $a_i \wedge b_i$ | $a_i \oplus b_i$ | $\left(a_i \wedge b_i\right) \wedge \left(a_i \oplus b_i\right)$ |
|---|---|---|---|---|
| 1 | 1 | 1 | 0 | 0 |
| 1 | 0 | 0 | 1 | 0 |
| 0 | 1 | 0 | 1 | 0 |
| 0 | 0 | 0 | 0 | 0 |

From table 4 it can be verified that

MCVT(x, y)= $\left(\left(a_n \wedge b_n\right) \wedge \left(a_n \oplus b_n\right), \left(a_{n-1} \wedge b_{n-1}\right) \wedge \left(a_{n-1} \oplus b_{n-1}\right), ................, \left(a_1 \wedge b_1\right) \wedge \left(a_1 \oplus b_1\right)\right)$

= (0, 0, 0, 0,……………………,0 )=0. Hence proved.

## 5. An Equivalence Relation is defined using the notion of CVT

Let A = $\left\{0,1,2,3, .................,2^n -1\right\}$ be a finite subset of $Z$ for some non-negative integer n and let R be a relation on A $\times$ A defined as (a, b) R (c, d) $\Leftrightarrow$ (a, b) and (c, d) requires equal number of iterations for getting their CVT=0 or XOR =0.

It can be easily verified that the relation R is reflexive, symmetric and transitive on the set A $\times$ A. Therefore R is an equivalence relation on A $\times$ A.

We have calculated the number of iterations required for the set of ordered pair in A $\times$ A where A={0,1,2……..,31} and constructed table 5 using a two step procedure as follows.

**Step 1** Write all the integers 0, 1, 2, 3,………,31 in ascending order in both, uppermost row and leftmost column of the table.

**Step2**. Compute number of iterations required for any ordered pair (a, b) to get either CVT=0 or XOR=0 and store it in the position (a, b).

From table 5, we have observed that
(1) The matrix is symmetric.
(2) If we consider the table as 4-quadrants, each quadrant is a symmetric matrix thentwo quadrants $1^{st}$ and $3^{rd}$ are same. Again if each quadrant is divided further into 4 smaller quadrants, then also the $1^{st}$ quadrant is same as the $3^{rd}$ quadrant. hence a self similar fractal behaviour is noticed in the table.

**(3)** In a block of size $(2^n-1)\times(2^n-1)$, there are no ordered pairs in the 2nd quadrant which transform into CVT=0 or XOR=0 in n-iterations.

Table 5: Showing the number of iterations required for either CVT=0 or XOR=0

The above relation R divides the set $\{0,1,2,3,................,2^n-1\}\times\{0,1,2,3,................,2^n-1\}$ into n disjoint equivalence classes.

For n=1, there is one equivalence class [(0 ,0)] ={(0,0) , (0 ,1) ,(1 ,0),(1,1)} and $\lVert 0,0\rVert = 4$

For n=2, there are two equivalence classes [(0, 0)], [(1, 3)] where

[(0, 0)] ={(0, 0), (0,1), (0,2), ( 0, 3), (1, 0), (1,1), (1, 2), (2, 0), (2,1), (2 ,2), (3, 0), (3, 3)}

[(1, 3)] = {(1, 3), (2,3), (3,1), ( 3, 2)} .

Here $\lVert 0,0\rVert = 12, \lVert 1,3\rVert = 4$

For n=3 , there are three equivalence classes [(0, 0)] , [(1, 3)] and [(1, 7)] .

$\lVert 0,0\rVert = 34, \lVert 1,3\rVert = 18, \lVert 1,7\rVert = 12$

For n=4, there are four equivalence classes [(0 ,0)] , [(1 ,3)] , [(1,7)] and [(1,15)] .

$\lVert 0,0\rVert = 96, \lVert 1,3\rVert = 78, \lVert 1,7\rVert = 58, \lVert 1,15\rVert = 24.$

For n=5, there are five equivalence classes [(0 ,0)] , [(1 ,3)] , [(1,7)] , [(1,15)] and [(1,31)] .

$\|[0,0]\|=274$, $\|[1,3]\|=306$, $\|[1,7]\|=263$, $\|[1,15]\|=133$, $\|[(1,31)]\|=48$

From above, we conclude that if we take a block of size $(2^n - 1) \times (2^n - 1)$, then

(1) number of ordered pairs for which CVT=0 or XOR=0 in n iterations is $3 \times 2^{n-1}$ for n=3, 4, 5,……

(2) number of ordered pairs for which CVT=0 or XOR=0 in one iterations is $3^n + (2^n - 1)$ for n=1, 2, 3, 4,…

## 6. Conclusion and Future Research Work

This paper provides some important properties of Carry Value Transformation (CVT) and Modified Carry Value Transformation (MCVT). We have proved that for any number system, the sum of any two non-negative integers is equal to the addition of their CVT and XOR values. We have also proved that in a successive addition of CVT and XOR of any two non-negative integers, the maximum number of iterations required to get either CVT=0 or XOR=0 is equal to the length of the bigger integer whereas the iterative process of the addition of MCVT and XOR requires exactly two steps for MCVT = 0. An equivalence relation is also defined in the set $Z \times Z$ which divides the CV table into disjoint equivalence classes.

In future we propose to study the following aspects
1. To investigate into the state transition diagrams (STDs) of different IVT.
2. To extend the domain of CVT from non-negative integers to real numbers and complex numbers.
3. To explore the behaviour of hybrid IVTs and their applications.
4. To explain the relationship of IVTs with Cellular Automata.

## Acknowledgement

The authors would like to acknowledge Prof. P. Pal Choudhury and Sk. Sarif Hassan, Indian Statistical Institute, Kolkata for motivating us to work further in the domain of CVT and IVT.